# CD2Alloy: Class Diagrams Analysis Using Alloy Revisited

Shahar Maoz*, Jan Oliver Ringert**, and Bernhard Rumpe

Software Engineering
RWTH Aachen University, Germany
http://www.se-rwth.de/

**Abstract.** We present CD2Alloy, a novel, powerful translation of UML class diagrams (CDs) to Alloy. Unlike existing translations, which are based on a shallow embedding strategy, and are thus limited to checking consistency and generating conforming object models of a single CD, and support a limited set of CD language features, CD2Alloy uses a deeper embedding strategy. Rather than mapping each CD construct to a semantically equivalent Alloy construct, CD2Alloy defines (some) CD constructs as new concepts within Alloy. This enables solving several analysis problems that involve more than one CD and could not be solved by earlier works, and supporting an extended list of CD language features. The ideas are implemented in a prototype Eclipse plug-in. The work advances the state-of-the-art in CD analysis, and can also be viewed as an interesting case study for the different possible translations of one modeling language to another, their strengths and weaknesses.

## 1 Introduction

The analysis of artifacts in one modeling language can, in many cases, be done using a semantics preserving translation to another language, and a reversed translation, back from the analysis results to the concepts of the first language. Often, more than one possible translation may be developed, and so, the definition of alternative translations, their implementation, and a comparative discussion on their strengths and weaknesses is worthwhile.

A UML class diagram (CD) can be analyzed using a translation to Alloy [1, 14]. The Alloy module is analyzed using a SAT solver, and the analysis result, an instance of the module, if any, can be translated back to the UML domain, as an object diagram. Existing translations [2, 3, 18, 22], however, are limited to this basic analysis of a single CD and are missing support for several CD language features, e.g., multiple inheritance and interface implementation, mainly because these features of CDs do not have direct, immediate counterparts in Alloy. In other words, they use a shallow embedding strategy.

* S. Maoz acknowledges support from a postdoctoral Minerva Fellowship, funded by the German Federal Ministry for Education and Research.
** J.O. Ringert is supported by the DFG GK/1298 AlgoSyn.

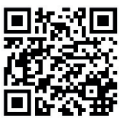

In this paper we present CD2Alloy, a new, alternative translation of UML CDs to Alloy, which is based on a deeper embedding strategy. Rather than mapping each CD construct to a semantically equivalent Alloy construct, our translation defines (some of) the CD constructs as new concepts within Alloy. For example, class inheritance is not mapped to its Alloy's counterpart — the `extends` keyword. Instead, it is defined using several of Alloy's language constructs — facts, functions, and predicates, whose semantics reflects the semantics of class inheritance in CDs.

The alternative translation we present has several advantages. First, it allows us to support more CD language features, in particular those features that do not have direct counterparts in Alloy, such as multiple inheritance and interface implementation. Second, significantly, it allows us to solve several analysis problems that go beyond the basic consistency check and instance generation tasks of a single CD, e.g., the analysis of the intersection of two CDs (i.e., generating common object models), the comparison of two CDs (checking if one is a refinement of the other), etc. These would have been very difficult, if not impossible, to support using existing translations from the literature.

Technically, as concrete languages we use the CD and object diagrams (OD) sublanguages of UML/P [20], a conceptually refined and simplified variant of UML designed for low-level design and implementation. Our semantics of CDs and ODs are based on [4, 8, 10] and are given in terms of sets of objects and relationships between these objects.

We define a transformation that takes one or more CDs and outputs an Alloy module. The Alloy module can then be analyzed with the Alloy Analyzer. Finally, using another transformation, instances of the Alloy module, if any, as found by the SAT solver connected to the Alloy Analyzer, are translated from Alloy back to ODs. The transformations are presented in Sect. 3. As mentioned above, the new translation allows us not only to support an extended list of CD language features but also to solve analysis problems that involve a number of CDs and could not have been solved before. We discuss the extension of the transformation from a single CD to multiple CDs, and some of the analysis problems we solve, in Sect. 4.

Our work is fully implemented in a prototype Eclipse plug-in we call CD2Alloy. CD2Alloy allows the engineer to edit a CD, to analyze it using Alloy, and to view the instances that the SAT solver finds, if any, back in the form of ODs. The analysis is fully automated, so the engineer need not see the generated Alloy code. We discuss the implementation in Sect. 5.

Sect. 2 gives brief background on the CD and OD languages and a short overview of Alloy. Sect. 3 describes our new translation from CDs to Alloy and back to ODs, side by side with the shallow translation described in [3]. Sect. 4 shows how the new transformation can be used to solve several analysis problems involving more than one CD. Sect. 5 presents the CD2Alloy plug-in. Sect. 6 summarises the comparison between the existing shallow translations and the new one, considering their strengths and weaknesses. Sect. 7 discusses related work and Sect. 8 concludes.

## 2 Preliminaries

### 2.1 Class and Object Diagrams

As concrete languages we use the CD and OD sublanguages of UML/P [20]. UML/P is a conceptually refined and simplified variant of UML designed for low-level design and implementation. Our semantics of CDs is based on [4, 8, 10] and is given in terms of sets of objects and relationships between these objects. More formally, the semantics is defined using three parts: (1) a definition of the syntactic domain, i.e., the syntax of the modeling language CD and its context conditions (we use MontiCore [15] for this), (2) a semantic domain, in our case, a subset of the System Model (see [4, 8]) OM, consisting of all finite object models, and (3) a mapping $sem : CD \to \mathcal{P}(OM)$, which relates each syntactically well-formed CD to a set of constructs in the semantic domain OM. A thorough and formal account of the semantics can be found in [8].

### 2.2 A brief overview of Alloy

Alloy [1, 14] is a textual modeling language based on relational first-order logic. An Alloy module consists of signature declarations, fields, facts and predicates. Each signature denotes a set of atoms, which are the basic entities in Alloy. Relations between two or more signatures are represented using fields and are interpreted as sets of tuples of atoms. Facts are statements that define constraints on the elements of the model. Predicates are parametrized constraints. A predicate can be included in other predicates or facts.

Alloy modules can be analyzed using Alloy Analyzer, a fully automated constraint solver. This is done by a translation of the module into a Boolean expression, which is analyzed by SAT solvers embedded within the Analyzer. The analysis is based on an exhaustive search for instances of the module, bounded by a user-specified scope, which limits the number of atoms for each signature in an instance of the system that the solver analyzes. The Analyzer can check for the validity of user-specified assertions: if an instance that violates the assertion is found within the given scope, the assertion is not valid, but if no instance is found, the assertion might be invalid in a larger scope. Used in the opposite way, the Analyzer can look for instances of user-specified predicates: if the predicate is satisfiable within the given scope, the Analyzer will find an instance that proves it, but if not, the predicate may be satisfiable in a larger scope. For a complete and detailed account of Alloy see [14].

## 3 The CD2Alloy Translation

We show our translation from CD to Alloy and from Alloy's instances back to ODs. We present these side by side with the shallow translation, described in [3], and focus on the key technical differences between the two (we chose to compare with [3] because it provides an implementation and appears to be the

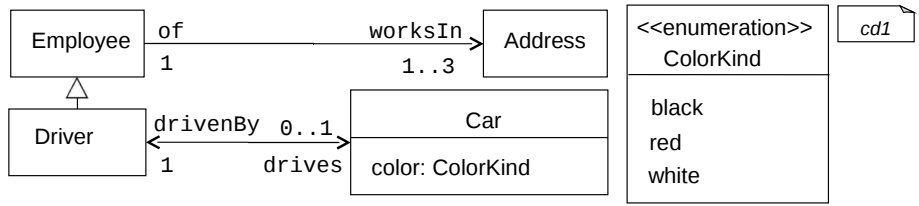

**Fig. 1.** $cd_1$, an example class diagram with classes, attributes, enumerations, associations with multiplicities, and inheritance

```
1  // Names of fields/associations in classes of the model
2  abstract sig FName {}
3
4  // Parent of all classes relating fields and values
5  abstract sig Obj { get: FName -> {Obj + Val + EnumVal} }
6
7  // Values of fields
8  abstract sig Val {}
9
10 // No values can exist on their own
11 fact { all v: Val | some f: FName | v in Obj.get[f] }
12
13 // Names of enum values in enums of the model
14 abstract sig EnumVal {}
15
16 // No enum values can exist on their own
17 fact { all v: EnumVal | some f: FName | v in Obj.get[f] }
```

**Listing 1.1.** Excerpt from the generic part of our translation: `FName`, `Obj`, `Val`, and `EnumVal` signatures and related facts

most advanced work of the shallow embedding approaches). The presentation uses the CD of Fig. 1 as a running example. We begin with an overview of our approach and continue with specific examples for various features. The complete translation will appear in an extended version of this paper.

### 3.1 From CD to Alloy

CD2Alloy takes a CD as input and generates an Alloy module. The module consists of a generic part (described below) and a CD specific part, which includes a predicate that describes the CD itself.

**The generic part** List. 1.1 shows the abstract signature `FName` used to represent association role names and attribute names for all classes in the module. The abstract signature `Obj` is the parent of all classes in the module; its `get` Alloy field relates it and an `FName` to instances of `Obj`, `Val`, and `EnumVal`. The

```
1  pred ObjAttrib[objs:set Obj,
2          fName:one FName, fType:set {Obj + Val + EnumVal}] {
3    objs.get[fName] in fType
4    all o: objs| one o.get[fName] }
5
6  pred ObjNoFName[objs:set Obj, fName:set FName] {
7    no objs.get[fName] }
8
9  pred ObjUAttrib[objs:set Obj,
10      fName:one FName, fType:set Obj, up: Int] {
11    objs.get[fName] in fType
12    all o: objs| (#o.get[fName] =< up) }
```

**Listing 1.2.** Excerpt from the generic part of our translation: parametrized predicates for the relations between objects and their fields, and for their multiplicities

abstract signature `Val`, which we use to represent all predefined types (i.e., primitive types and other types that are not defined as classes in the CD). Values of enumeration types are represented using the signature `EnumVal`. Enumeration values and primitive values should only appear in an instance if they are referenced by an object (as specified by the facts in line 11 and line 17).

List. 1.2 shows some of the generic, parametrized predicates responsible for specifying the relation between objects and fields: `ObjAttrib` limits `objs.get[fName]` to the correct field's type and ensures that there is exactly one object, value, or enumeration value related to the field name by the `get` relation; `ObjFNames` is used to ensure objects do not have field names other than the ones stated in the CD. List. 1.2 also shows one of the generic predicates responsible for specifying association multiplicities: `ObjUAttrib` provides an upper bound for the number of objects in the set represented by the `get` relation for a specified role name.

All the above are generic, that is, they are common to all generated modules, independent of the input CD at hand. We now move to the parts that are specific to the input CD, and present specific examples of various features.

**Classes and attributes** Consider a fragment of the CD shown in Fig. 1 consisting of only the class `Car` and its `color` attribute. With the transformation of [3], this fragment translates to the Alloy code shown in List. 1.3. In our transformation, this fragment translates to the Alloy code shown in List. 1.4.

**Associations** We continue with associations, where directions and multiplicity ranges need to be expressed. To support bidirectional associations and custom multiplicity ranges in the shallow translation of [3], engineers are required to manually write the specific OCL constraints that characterize these features, because Alloy does not have a direct counterpart to the concept of association and its signature field definition does not have explicit built-in support for cardinalities. In our work, however, the semantics of bidirectionality and custom

```
1 sig Car{color:one ColorKind}
2 abstract sig ColorKind{}
3 one sig black extends ColorKind{}
4 one sig red extends ColorKind{}
5 one sig white extends ColorKind{}
```
**Listing 1.3.** Car with color in the translation of [3]

```
1 one sig color extends FName {}
2
3 lone sig enum_ColorKind_black extends EnumVal {}
4 lone sig enum_ColorKind_red extends EnumVal {}
5 lone sig enum_ColorKind_white extends EnumVal {}
6
7 sig Car extends Obj {}
8
9 fun ColorKindEnum: set EnumVal  {
10   enum_ColorKind_black +
11   enum_ColorKind_red +
12   enum_ColorKind_white }
13
14 pred cd {
15   ObjAttrib[Car, color, ColorKindEnum]
16   ObjFNames[Car, color] }
```
**Listing 1.4.** Car with color in our translation

multiplicity ranges is part of the translation itself: there is no need for manual OCL writing to express these standard concepts.

For example, consider a fragment of the CD shown in Fig. 1 consisting of only `Employee` and `Address`, and the association `worksIn` between them. With the transformation of [3], this fragment translates to the Alloy code shown in List. 1.5.[1] It is translated in our transformation to the code shown in List. 1.6.

**Single inheritance, interfaces, and multiple inheritance** We now extend the examples above with inheritance. We show how the two translations handle single inheritance and how our translation can also support interfaces and multiple inheritance.

The translation of [3] takes advantage of Alloy's built-in support for inheritance, and thus directly maps CD class inheritance to Alloy's `extends` keyword. In our translation, in contrast, the semantics of inheritance, that is, the meaning of the 'is-a' relation in terms of inclusion between sets, is explicitly expressed using sub class functions. The inheritance hierarchy is flattened and then rebuilt:

---

[1] According to our experience, UML2Alloy tool of [3], version 0.5.2, does not support such multiplicity ranges without the manual addition of OCL expressions. The above code shows how the translation of UML2Alloy could have handled multiplicity ranges if it supported this feature.

```
1 sig Address{}
2 sig Employee{worksIn:set Address}
3 fact Asso_Employee_of_worksIn_Address { Employee <:
4   worksIn in ( Employee) one->set ( Address) }
5 fact AssoCustom_Employee_of_worksIn_Address {
6   all var:Employee| #var.worksIn =< 3 && #var.worksIn >= 1}
```

**Listing 1.5.** `Employee` works in `Address` in the translation of [3]

```
1 one sig of,worksIn extends FName {}
2
3 sig Address,Employee extends Obj {}
4
5 fun AddressSubs: set Obj {Address}
6 fun EmployeeSubs: set Obj {Employee}
7
8 pred cd {
9    ObjFNames[Address, of]
10   ObjFNames[Employee, worksIn]
11   ObjLUAttrib[EmployeeSubs, worksIn, AddressSubs, 1,3]
12   ObjLUAttrib[AddressSubs, of, EmployeeSubs, 1,1] }
```

**Listing 1.6.** `Employee` works in `Address` in our translation

in particular, as part of flattening, the complete list of attributes and associations of each class is collected from all its super classes. The sub class functions define the set of sub classes of each class.

Listings 1.7 and 1.8 show the parts related to inheritance in the Alloy code for the example CD of Fig. 1 in the two translations. Note the `EmployeeSubs` function in line 8 of List. 1.8 which returns the set of sub classes of `Employee`.

Similar functions are used to support interfaces. For every interface we define a function which returns all classes implementing it.

Significantly, consider a different CD where the class `Driver` does not inherit `Employee`, but where a new class `Chauffeur` inherits both `Driver` and `Employee`. This multiple inheritance setup is not supported by shallow translations like the one of [3] but it is supported by our translation (see List. 1.9). The use of functions provides the flexibility required to support multiple inheritance.

**Composition** Our translation supports a whole/part composition relation. Composition is not supported by shallow translations like the one of [3] because CD's composition has no direct counterpart construct in Alloy.

The semantics of composition requires that a part cannot exist without a whole and that it belongs to exactly one whole. The predicate for composition is shown in List. 1.10. This predicate can be used, e.g., to specify a composition relation between `Employee` and `Address`, by adding the statement `Composition [EmployeeSubs , worksIn , AddressSubs]` to the CD predicate.

```
1  sig Address{}
2  sig Employee{worksIn:set Address}
3  sig Car{drivenBy:one Driver}
4  sig Driver extends Employee{drives:one Car}
```
**Listing 1.7.** `Driver` inherits from `Employee` in the translation of [3]

```
1  one sig drivenBy,of,worksIn,drives extends FName {}
2
3  sig Driver,Car,Address,Employee extends Obj {}
4
5  fun DriverSubs: set Obj {Driver}
6  fun CarSubs: set Obj {Car}
7  fun AddressSubs: set Obj {Address}
8  fun EmployeeSubs: set Obj {Employee + Driver}
9
10 pred cd {
11    ObjFNames[Driver, drives]
12    ObjFNames[Car, drivenBy]
13    ObjFNames[Address, of]
14    ObjFNames[Employee, worksIn]
15    ObjLUAttrib[EmployeeSubs, worksIn, AddressSubs, 1,3]
16    ObjLUAttrib[AddressSubs, of, EmployeeSubs, 1,1]
17    BidiAssoc[DriverSubs, drives, CarSubs, drivenBy]
18    ObjLUAttrib[CarSubs, drivenBy, DriverSubs, 1,1]
19    ObjLUAttrib[DriverSubs, drives, CarSubs, 0,1] }
```
**Listing 1.8.** `Driver` inherits from `Employee` in our translation

### 3.2 Back to UML object diagrams

Finally, we discuss the translation back from Alloy instances to UML ODs. In the translation presented in [3], the translation of an Alloy instance back to a UML OD is an immediate one to one mapping, which, according to [22], can be automatically computed from the first translation. Each atom is transformed, directly, into a UML object.

In contrast, in our translation, object instances are constructed only for the atoms in the Alloy instance that are instances of `Obj`; for each of these, attributes and their values are computed from the instances of their `get` relation (see line 5 of List. 1.1). More specifically, an Alloy instance that is found for a module generated by our translation may also include atoms that do not correspond to objects in the object model it represents, e.g., field names and enumeration values. Thus, these should not be translated to objects in the translation back to UML ODs. This makes our transformation from Alloy instances back to UML somewhat complicated. The resulting OD is a valid UML object diagram that indeed describes an instance of the original CD in terms of UML semantics.

```
1 one sig drivenBy,of,worksIn,drives extends FName {}
2
3 sig Driver,Car,Address,Employee,Chauffeur extends Obj {}
4
5 fun DriverSubs: set Obj {Driver + Chauffeur}
6 fun CarSubs: set Obj {Car}
7 fun AddressSubs: set Obj {Address}
8 fun EmployeeSubs: set Obj {Employee + Chauffeur}
9 fun ChauffeurSubs: set Obj {Chauffeur}
10
11 pred cd {
12    ObjFNames[Driver, drives]
13    ObjFNames[Car, drivenBy]
14    ObjFNames[Address, of]
15    ObjFNames[Employee, worksIn]
16    ObjFNames[Chauffeur, worksIn + drives]
17    ObjLUAttrib[EmployeeSubs, worksIn, AddressSubs, 1,3]
18    ObjLUAttrib[AddressSubs, of, EmployeeSubs, 1,1]
19    BidiAssoc[DriverSubs, drives, CarSubs, drivenBy]
20    ObjLUAttrib[CarSubs, drivenBy, DriverSubs, 1,1]
21    ObjLUAttrib[DriverSubs, drives, CarSubs, 0,1] }
```

**Listing 1.9.** Multiple inheritance: `Chauffeur` inherits both `Driver` and `Employee`, in our translation (note the functions in lines 5 and 8)

```
1 pred Composition[wholes: set Obj,
2          rName: some FName, parts: set Obj] {
3   all p: parts | #{w: wholes, r: rName | p in w.get[r]}=1 }
```

**Listing 1.10.** A predicate for whole/part composition relation

## 4 Multiple CD Analysis

In addition to supporting an extended list of CD language features, the new translation allows us to solve several analysis problems beyond the basic, single CD consistency and instance generation. We show how the translation described in the previous section is generalized to support multiple CDs and continue to present its application to two analysis problems.

### 4.1 Handling multiple CDs

To handle multiple CDs in one Alloy module, we define signatures for the union of classes from all input CDs, and divide the CD specific functions (sub class functions, enumeration value functions) between the CDs by adding a suffix CD$i$ to all functions generated for the $i$-th CD. Moreover, instead of creating a single predicate cd, we generate several predicates, cd1, cd2, etc., one for each of the input CDs. Each predicate cd$i$ uses the functions with suffix CD$i$ and defines constraints to not include any objects of classes not in $cd_i$. This is necessary because the predicate is interpreted as part of the module, which contains signatures representing classes from other CDs too.

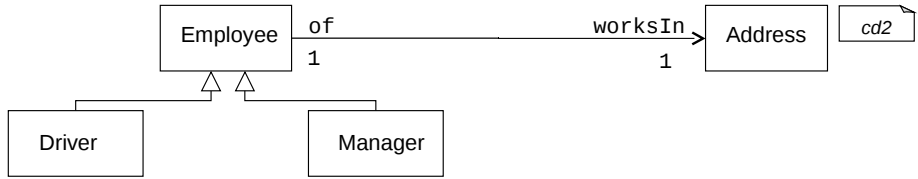

**Fig. 2.** $cd_2$, an example CD for the computation of intersection with $cd_1$ (see Sect. 4.2)

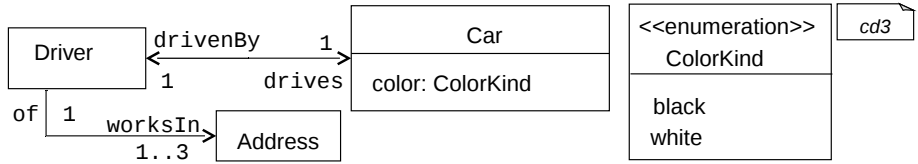

**Fig. 3.** $cd_3$, an example CD that refines $cd_1$ (see Sect. 4.3)

### 4.2 Example analysis problem: intersection

As one example application, we show how to use our translation to check the intersection of the semantics of two (or more) CDs. Recall the CD shown in Fig. 1 and consider a second CD, as shown in Fig. 2. Is there a system that satisfies both CDs, i.e, do the two CDs have common object model instances?

To answer this question using our translation we ask Alloy to find instances of the predicate `cd1 and cd2`. If any exist, we know that the intersection of the two CDs semantics is not empty. List. 1.11 shows snippets from the Alloy module corresponding to checking the intersection of the two CDs in our example, $cd_1$ of Fig. 1 and $cd_2$ of Fig. 2. Analyzing the predicate `cd1 and cd2` reveals that their intersection is not empty: for example, an object model consisting of two drivers, each with one address, is an instance of both CDs. This object model can be found when executing the analyzer on the predicate `cd1 and cd2` with our translation of the two CDs.

### 4.3 Example analysis problem: refinement

The above technique can be easily generalized to solve the consistency of any Boolean expression over a set of CDs. For example, an analysis of the predicate `cd1 and not cd2` would find instances of the first CD that are not instances of the second, if any.

So, as a second analysis problem, we show that our translation can be used to check for refinement relations between CDs: If the predicate `cd1 and not cd2` is inconsistent (has no instances) and the predicate `cd2 and not cd1` is consistent

```
1  // signatures for the union of classes from the two CDs
2  // ...
3
4  // functions with CD# suffix
5  fun DriverSubsCD1: set Obj {Driver}
6  fun EmployeeSubsCD1: set Obj {Employee + Driver}
7  fun DriverSubsCD2: set Obj {Driver}
8  fun EmployeeSubsCD2: set Obj {Employee + Driver + Manager}
9  // more functions...
10
11 pred cd1 {
12   // use functions with suffix CD1
13   // ...
14   no Manager }
15
16 pred cd2 {
17   // use functions with suffix CD2
18   // ...
19   no Car }
20
21 run {cd1 and cd2} for 10
```

**Listing 1.11.** Checking the intersection of $cd_1$ and $cd_2$ using our translation

(has instances), we can conclude that all instances of $cd_1$ are also instances of $cd_2$ (but not the other way around), namely, that $cd_1$ is a (strict) refinement of $cd_2$. As a concrete example, recall CD $cd_1$ of Fig. 1 and consider $cd_3$ shown in Fig. 3. Analyzing `cd1 and not cd3` and `cd3 and not cd1` reveals that all instances of $cd_3$ are indeed instances of $cd_1$, but not the other way around. Thus, the analysis shows that $cd_3$ is a refinement of $cd_1$.

To the best of our understanding, such analyses are not possible in existing translations.

## 5 Implementation: The CD2Alloy Plug-In

Our work is implemented in a prototype Eclipse plug-in called CD2Alloy. The input for the implementation is a UML/P CD, textually specified using Monti-Core grammar and generated Eclipse editor [15]. The transformation to Alloy is implemented using templates written in FreeMarker [11] and the execution of the generated module's run commands is done using Alloy's APIs [1]. The analysis is fully automated so the engineer does not need to see the generated Alloy code (viewing the generated Alloy code is optional).

CD2Alloy allows the engineer to edit a CD, to analyze it using Alloy, and to view the instances that the SAT solver finds back in the form of ODs. The plug-in, together with relevant documentation and examples, is available from [7]. We encourage the interested reader to try it out.

## 6   Discussion

The analysis of artifacts in one modeling language can, in many cases, be done using a semantics preserving translation to another language (and a reversed translation, back from the analysis results to the concepts of the first language). Often, more than one possible translation may be developed, and so, a comparative discussion on the characteristics of such translations and their implementation is worthwhile. Our work may be viewed as an interesting case study example of the differences between two different translations, their strengths and weaknesses, in particular when they are used in the context of mechanized analysis (rather than, say, in the context of a pure theoretical definition of a semantics).

Strengths of the translations of [3, 18], and other shallow translations, are readability and relatively simple definition and implementation. The translation of each class requires only a local analysis and the resulting module syntax is linear in the size of the input CD. Weaknesses are the limited list of language features and possible potential analyses supported; these translations do not take full advantage of the expressive power of Alloy to cover the rich features of CDs.

Strengths of the new translation are twofold. First, the powerful possible analyses, such as refinement checking, mounting to evaluating any Boolean expression over CDs. Second, the extended list of features, including multiple inheritance and interface implementation, which have no direct counterparts in Alloy and are thus handled using a deep embedding strategy. Supporting these is important not only for theoretical coverage of language features but also because many CDs in the real world do make significant use of them.

One weakness of the new translation is that it is more difficult to read and understand, because there is no direct explicit mapping between the syntax of the generated module and the syntax of the CD. However, readability may be not so important in our context because the analysis is fully automated and the results are translated back to the UML domain.

Another weakness of the new translation is that it is harder to implement and more computationally complex: the flattening of the inheritance hierarchy requires a global analysis of the CD and in the worst case its reconstruction using functions may result in a module whose size is quadratic in the size of the input CD. This leads to a larger formula for the SAT solver used by Alloy.

As an example for the differences in computation complexity and performance, according to our experience, checking the consistency of the CD of Fig. 1 by generating an instance, with Alloy scope 3, using UML2Alloy (the tool described in [2, 3]), resulted in a SAT formula of 618 variables and 1025 clauses. Using our new translation, CD2Alloy, the same problem resulted in a formula of 3354 variables and 5627 clauses. SAT solving time increased too, from 6 to 14 milliseconds (using SAT4J, on a Dell Latitude E6500 laptop running Windows 7). Note, however, that the use of the same scope in this comparison may be misleading: in the translation of [3], the scope defines the maximal number of objects per class in the instance, while in CD2Alloy, the scope defines the maximal number of objects in the instance.

To conclude, our work clearly demonstrates the tradeoff between the readability and intuitiveness of a simple shallow translation on the one hand and the expressiveness of a deeper translation on the other hand. The choice of translation to use depends on the specific needs of the applications at hand.

Finally, it is important to note that all existing translations, our new translation, and any other analysis performed with the Alloy Analyzer, are subject to a scope, which limits the number of atoms per signature (see [14]). In particular, it may be the case that a predicate does not hold in one scope but holds in a larger one. For an unbounded analysis one would need a translation of CDs to other formalisms, e.g., to enable the use of theorem provers, giving up full-automation, as in [5, 13].

## 7 Related Work

In [2, 3], the authors present a tool called UML2Alloy and provide a detailed discussion of the challenges of transforming CDs and OCL expressions into Alloy. One strength of this work is that the transformation used is defined and implemented using an MDA technique, that is, by formally defining a metamodel for CDs, a metamodel for Alloy, and transformation rules between the two. However, the shallow nature of the transformation between these metamodels limits the set of UML CD features that the work supports. For example, as multiple inheritance cannot be directly represented in Alloy, it is not supported by this work and is explicitly disallowed by the related profile (see [3, pp. 75]). Following an in-depth discussion of the differences between the languages, the authors of [3] conclude that "Because of these differences, model transformation from UML to Alloy has proved to be very challenging." [3, pp. 70]. Indeed, our work proposes to address this challenge by means of a deeper embedding strategy that bridges some of the differences between the languages: it takes advantage of Alloy's own expressive power to represent CD concepts that cannot be mapped directly to semantically equivalent concepts in Alloy.

A related work by some of the same authors [22] uses the same MDA approach, transformation, and tool, and adds a round trip transformation, from Alloy's instances back to UML ODs, implemented in QVT [19]. As we have shown in Sect. 3, our work supports a backward translation which results in correct ODs, i.e., ones which represent valid instances of the original CD according to the UML semantics. Supporting a translation back to the UML space is of course critical to the usefulness of the entire approach in practice.

In [18], the authors suggest to analyze CDs with Alloy, using a shallow embedding similar to [3]. This work does not present an implementation.

In [9], the authors use Alloy to formalize UML package merge. The work models a fragment of the UML metamodel in Alloy, in order to check various properties of package merge. Unlike our approach, this work does not present a generic transformation to Alloy. Analyses of multiple models are not discussed.

In [21], Sen presents a translation of the UML metamodel to Alloy, formalized and implemented in Kermeta. Similar to our work, this translation is not shallow and handles an extended list of CD features such as multiple inheritance and composition. Different from our work, it does not support analyses of multiple input models such as checking refinement and intersection.

UMLtoCSP [6] verifies UML/OCL models by a translation to a constraint satisfaction problem, solved using a constraint solver within a user-defined bounded search space. The tool checks for various kinds of satisfiability (and other analysis problems), and can generate an example instance (object model). Our work has similar strengths and weaknesses: the analysis is fully automated but is conducted in a bounded scope. We do not know whether UMLtoCSP supports multiple inheritance. It may be possible to extend UMLtoCSP to check for Boolean expressions over CDs, as supported by our work.

The USE tool [12] supports the analysis of CDs and related OCL invariants, checking, e.g., the consistency of a single CD, the independence of an OCL invariant, etc. A more recent work by the same group [23] reports on analyzing UML/OCL models directly using a SAT solver. To the best of our knowledge, applications such as checking refinement between two CDs are not available in [12, 23], but it may be possible to extend these works to support such applications.

Finally, in recent work [16] we have defined a semantic differencing operator for CDs (used for semantic model comparison in the context of model evolution), which we have implemented using a translation to Alloy, similar to the one presented here. This work takes two CDs as input and outputs an Alloy module whose instances represent *diff witnesses*, object models in the semantics of one CD that are not in the semantics of the other. Also, in another recent work [17] we use a variant of the translation presented here and extend it to support semantic variability in CD/OD consistency analysis. This work takes three artifacts as input: a CD, an OD, and a feature configuration, which specifies choices over a set of semantic variability points; the analysis is semantically configured and its results change according to the semantics induced by the selected feature configuration. These works are additional examples for the kinds of analyses enabled by our translation.

## 8  Conclusion

We have presented CD2Alloy, a translation from UML CDs to Alloy, which is deeper and qualitatively different than previously suggested translations. Our translation takes advantage of Alloy's expressive power and advances the state-of-the-art in CD analysis in several ways: (1) support for more CD language features and (2) support for solving additional analysis problems concerning multiple CDs. The ideas are implemented in a prototype Eclipse plug-in and demonstrated with running examples.

Future work includes the investigation of additional possible embeddings of fragments of UML into Alloy, in order to support additional language features and analyses, for example, constrained generalization sets.